\documentclass[aps,amsmath,prb,twocolumn,showpacs]{revtex4}
\usepackage{epsfig}
\usepackage{amsfonts}
\usepackage{amssymb}
\usepackage{afterpage}
\def\bibhookdar#1#2{\expandafter\xdef\csname bibhookdar:#1\endcsname{#2}}
\def\bibhookdarexp#1{\ifx\expandafter\csname bibhookdar:#1\endcsname%
\undefined\relax\else\expandafter\csname bibhookdar:#1\endcsname\fi}
\newwrite\figcapwrite\immediate\openout\figcapwrite\jobname-figcap.aux
\long\def\gencap#1#2{\immediate\immediate\write#2{\expandafter\noexpand#1\par}}
\long\def\figcap#1{%
\gencap{Figure \thefigure. \expandafter\noexpand#1}{\figcapwrite}}
\let\hackcaption\caption
\begin{document}
\clubpenalty5000\widowpenalty5000
\title{Thermoelectric Modeling of the Non-Ohmic Differential
Conductance in a Tunnel Junction containing a Pinhole}
\author{Z.-S. Zhang}
\author{D.A. Rabson}
\thanks{Corresponding author}
\affiliation{Department of Physics, University of South Florida,\\
Tampa, FL 33620, USA}
\def\ugly#1{\textbf{[#1]}}

\begin{abstract}
To test the quality of a tunnel junction,
one sometimes fits the bias-dependent differential conductance
to a theoretical model, such as Simmons's formula.  Recent
experimental work by {\AA}kerman and collaborators,
however, has demonstrated that a good
fit does not necessarily imply a good junction.  Modeling the
electrical and thermal properties of a tunnel junction containing
a pinhole, we extract an effective barrier height and effective barrier
width even when
as much as 88\% of the current flows through the pinhole short
rather than tunneling.  A good fit of differential conductance
to a tunneling form therefore cannot rule out pinhole defects
in normal-metal or magnetic tunnel junctions.
\end{abstract}


\maketitle

\section{Introduction}
With recent rapid progress toward incorporating tunnel junctions
into practical devices, such as magnetic sensors and MRAM,
\cite{Parkin99,Daughton,Reohr02,Motorola03}
it has become increasingly important to assure junction
quality.  Pinhole shorts through the insulating layer of a
nominal tunnel junction constitute a possible failure mode.
As early as the 1960s and 1970s, a set of criteria emerged, attributed
to Rowell (who also credits Giaever),\cite{Rowell69}
for determining whether a particular device
was a good tunnel junction.  Only a few apply when neither
metal junction superconducts.  One is the exponential dependence
of the resistance on the thickness of the insulating layer over a series
of devices with increasing thickness;
however, a simple model shows that classical pinholes
can mimic this
signature of tunneling.\cite{Rabson01}

Another, more commonly
used, criterion is a non-Ohmic differential
conductance.\cite{Moodera99,Wang01}
Simmons\cite{simmons} and
Brinkman, Dynes, and Rowell\cite{Brinkman} calculated the
tunneling current $I$ through idealized junctions as functions of
bias $V$, the height of the energy barrier, and the thickness of the insulator;
when differential
conductance $dI/dV$ shows positive curvature, can be fit to one of these
forms, and yields ``reasonable'' values for the barrier height and thickness,
comparable to independent measurements, a good junction
is sometimes presumed.
{\AA}kerman and collaborators
have demonstrated experimentally that this criterion alone cannot
establish the quality of a junction.\cite{Jonsson00,Akerman02}
They made a set of junctions with
one normal and one superconducting lead.  Above the superconducting-transition
temperature, all
fit the tunneling models well.  Below the superconducting transition,
in some of the junctions the current was suppressed inside the
gap, as would be expected for tunneling from a normal metal through
an intact insulator into a superconductor.  Other junctions, however,
instead showed the {\it enhancement\/} of current inside the gap
typical of Andreev reflection, indicating essentially direct
superconductor-metal contact.\cite{BTK82,BTK83}

We now demonstrate, in a simple model, how a tunnel junction
with a pinhole short might reproduce the non-superconducting result,
finding that
as much as 88\% of the current might flow through the short
and still leave enough tunneling to give a good fit to the Simmons
model.

A sample known to harbor a pinhole may show a differential conductance
with either positive
or negative curvature,\cite{Schmalhorst} the latter a heating
effect: a pinhole will dissipate more heat at greater bias,
becoming hotter, and, assuming it acts as a metal, conduct
less.
By contrast,
the conductance of a tunneling channel increases with bias,
as carriers see an effectively narrower trapezoidal
barrier.\cite{Duke69,Quantum}
The pinhole and the tunneling channels, therefore, have opposite
effects on the curvature, the one tending to make the conductance curve
down, the other making it curve upward.  For parameters based on
an experimental geometry, we find that the large positive curvature
of a relatively small tunnel current can overwhelm the weakly negative
curvature of the conductance through the pinhole short, still yielding
an excellent fit to the Simmons form with apparent
barrier width and height varying from the true width and height
by factors of 2--3.

Although two of the Rowell criteria are found unreliable,
the temperature dependence of conductance still appears to distinguish
good junctions from shorted ones.\cite{Jonsson00,Akerman01,Rudiger01,Akerman02}
Recently, we have proposed an additional test, using only electrical
measurements at a single temperature, that may serve both to diagnose
the presence of a pinhole and to determine its location.\cite{Zhang03}

\section{Model and Computation}
We build a three-dimensional network, 
each node of which is defined by  
material, position, thickness, and length.
The voltage and temperature at each node are computed iteratively,
in turn determining the local electrical resistivity
and thermal conductivity.
We extract our dimensions from the experimental geometry in reference
\onlinecite{Jonsson00}:
the side of the tunnel junction is 50$\mu$m;
the bottom electrode is Nb of thickness 80$\,$nm.
On top of the Nb layer are
an Al layer 8.5$\,$nm thick and a
2.0$\,$nm AlO$_{\textrm{x}}$ barrier;
then Fe (50$\,$nm) is deposited on top.
We set the barrier height at 0.5$\,$eV.
We model the pinhole as a metallic (Al) right square prism
at the center of the tunnel junction.
We alternate
calculations of the voltage and the temperature at each
node until reaching a steady state before
measuring
the effective resistance of the junction.

The electrical boundary conditions fix one edge of the bottom
metallic layer at ground and one edge of the top metallic
layer (rotated $90^\circ$ relative to the grounded edge)
at some positive voltage (the ``input voltage'').  The measured bias is
the difference between the average voltages on the two edges
opposite the fixed edges.  (This simulates a four-terminal
measurement of the sort we have previously
modeled in Reference \onlinecite{Zhang03}.  The measured bias
is generally somewhat smaller than the input voltage.)
We hold the bottom metallic layer in equilibrium with a fixed
heat bath; all heat generated in the junction must leave through
this surface.  All other surfaces are assumed perfect thermal insulators.
(Radiation losses are ignored.)


\begin{figure}[t]
\centerline{\epsfxsize0.8\hsize\epsfbox{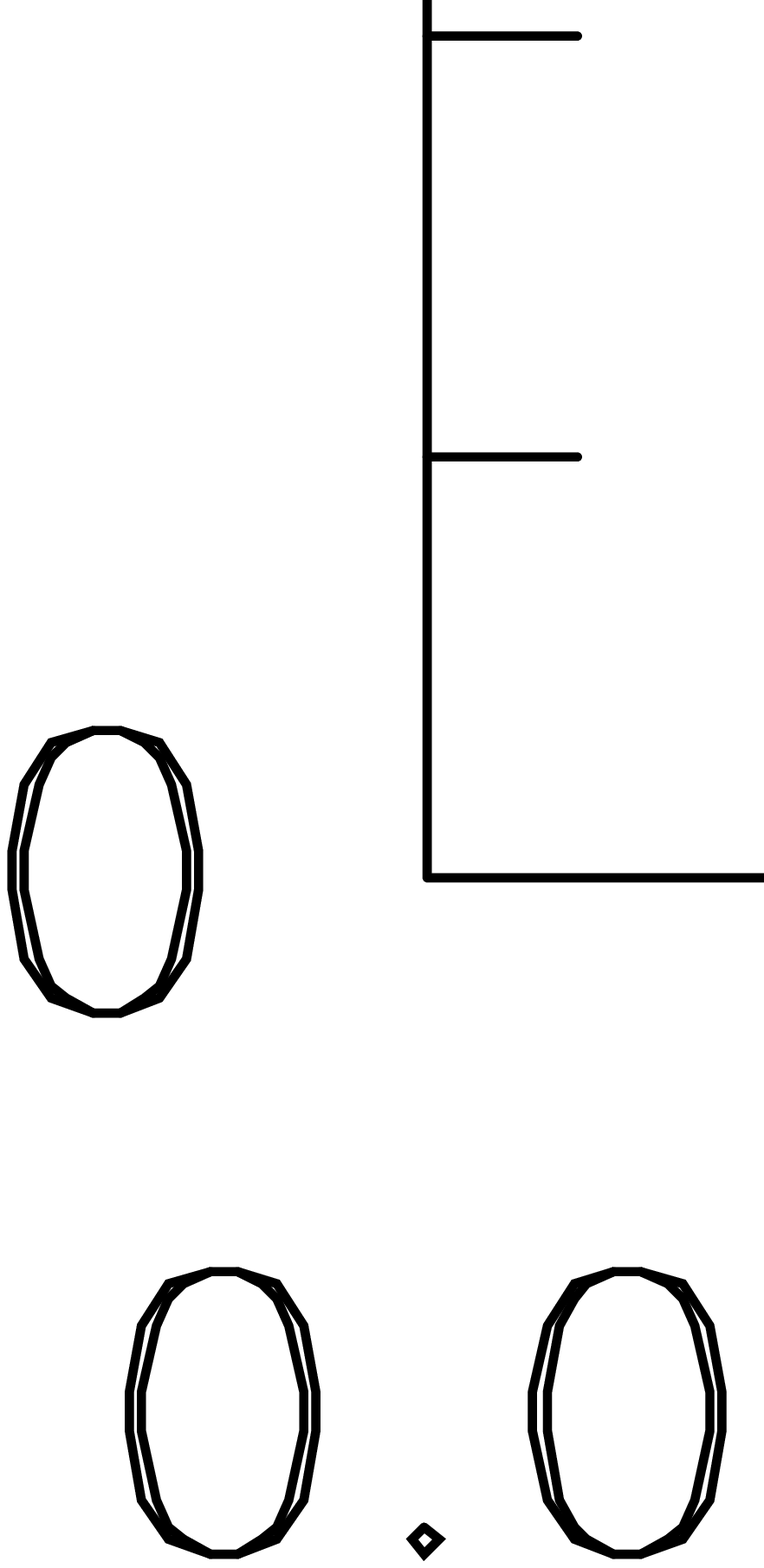}}
\hackcaption{Normalized differential conductance against bias. The dotted line
plots differential conductance absent any pinhole; the solid
lines trace conductance as
the pinhole side
changes (top to bottom) from 0.1$\,$nm to 1.5$\,$nm in increments of 
0.1$\,$nm.  The separatrix between negative and positive curvature
corresponds to a pinhole side between 1.0$\,$nm and 1.1$\,$nm.
The bottom-layer temperature is 77K.}
\label{d_c}
\end{figure}

\begin{figure}[b]
\centerline{\epsfxsize0.8\hsize\epsfbox{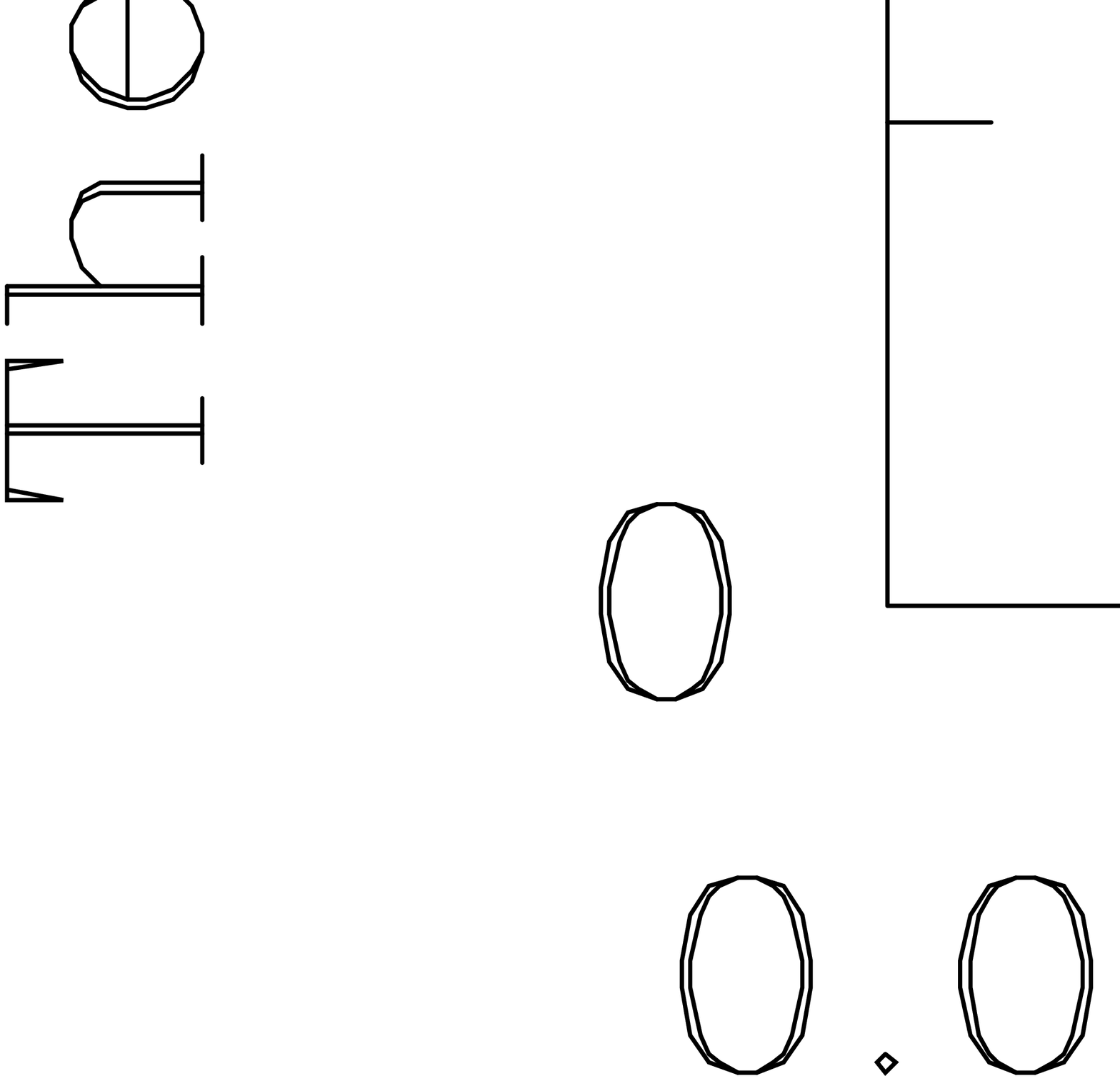}}
\hackcaption{The effective thickness of the barrier against  
the side of the pinhole. 
The true barrier thickness is 2.0$\,$nm.
The bottom-layer temperature is 77K.}
\label{e_t}
\end{figure}

\begin{figure}[t]
\centerline{\epsfxsize0.8\hsize\epsfbox{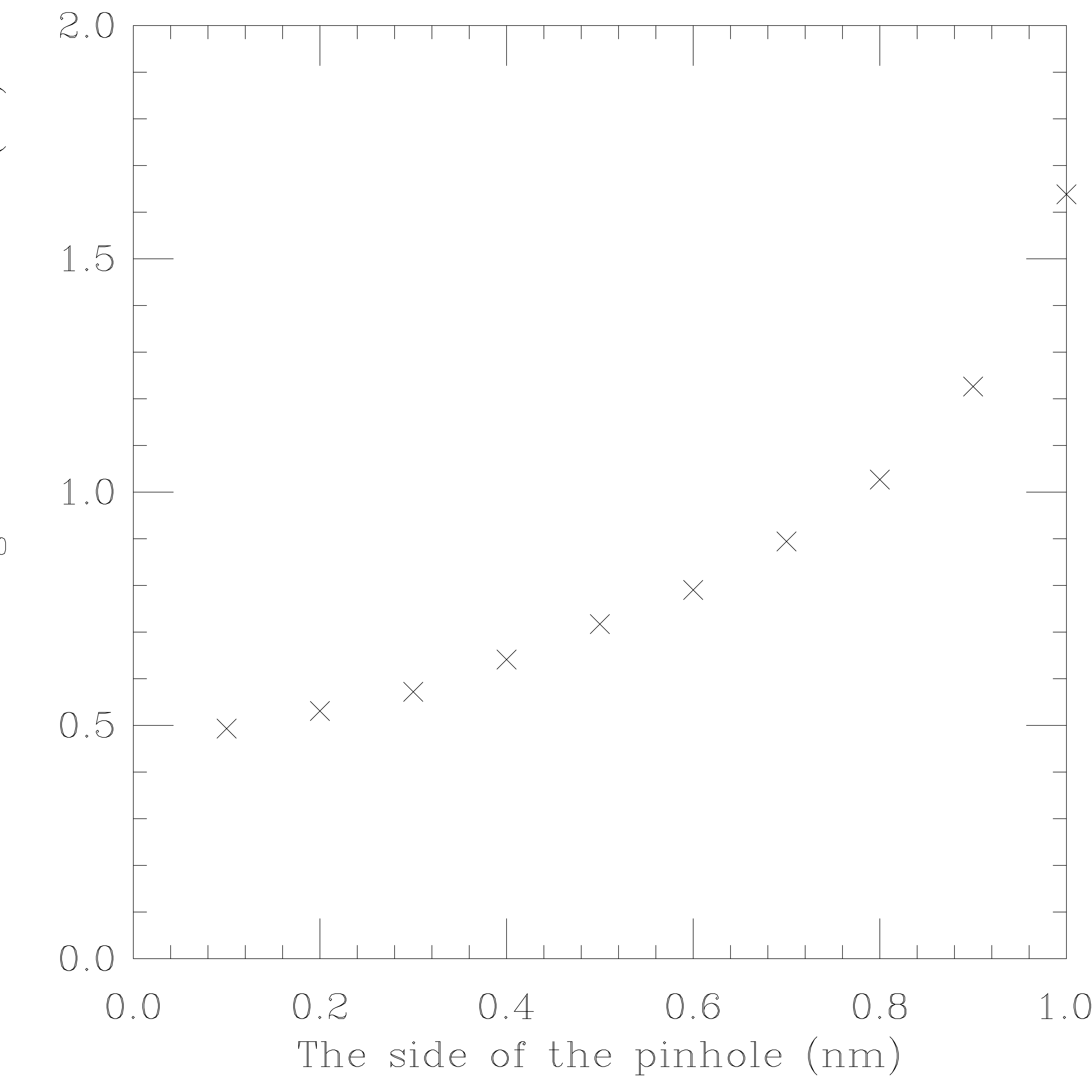}}
\hackcaption{The effective height of the barrier against 
the side of the pinhole. 
The true barrier height is 0.5$\,$eV. The 
temperature is 77K. }
\label{e_h}
\end{figure}

\begin{figure}[b]
\centerline{\epsfxsize0.8\hsize\epsfbox{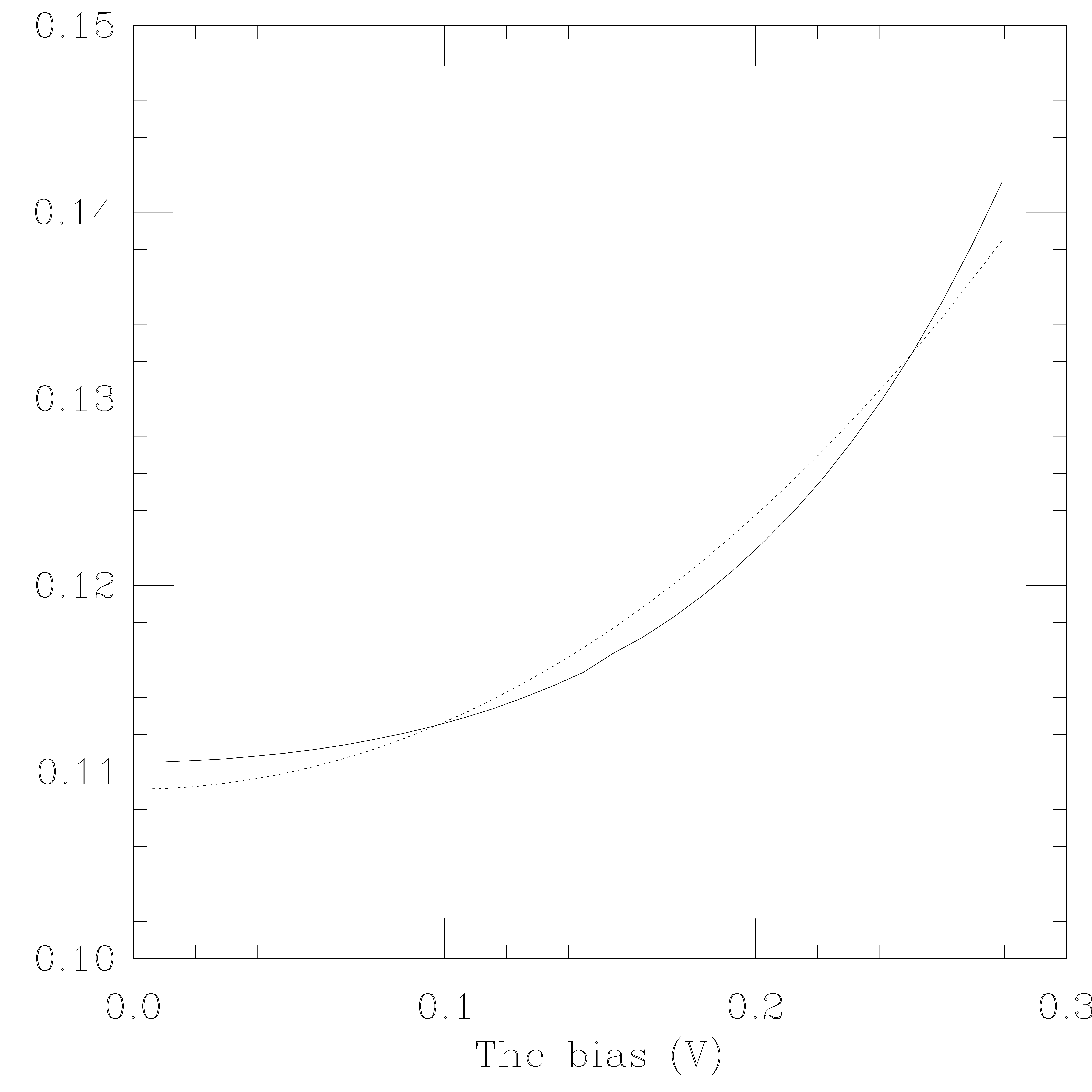}}
\hackcaption{Even in the presence of a large pinhole (0.9$\,$nm on a side),
the curvature of differential conductance can be fit to the
form of Simmons.
The solid line plots the simulation,
while the dotted line plots Simmons's formula for the
best-fit barrier height and thickness.
The bottom-layer temperature
is 77K.}
\label{t_c}
\end{figure}

\begin{figure}[t]
\centerline{\epsfxsize0.8\hsize\epsfbox{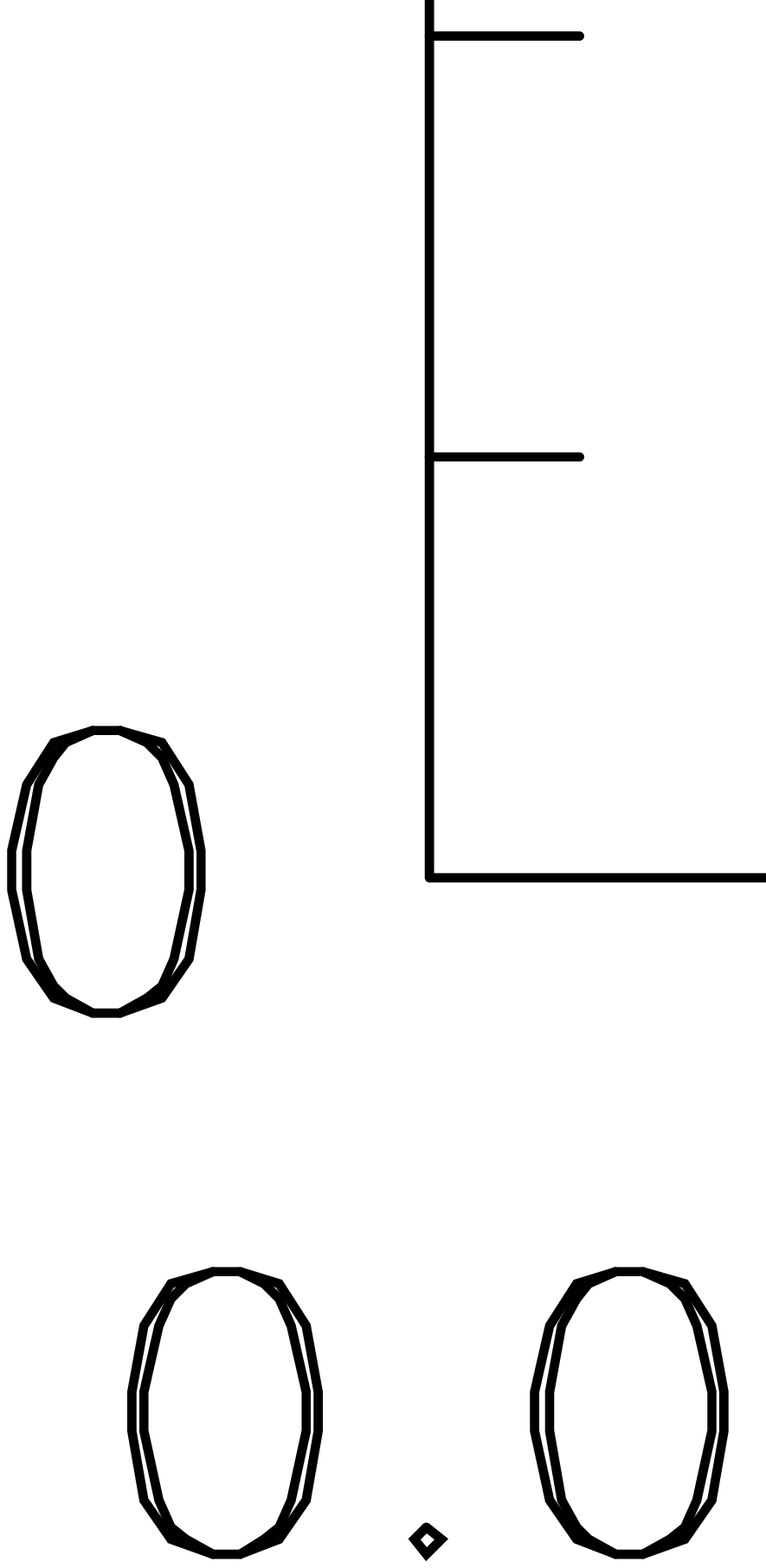}}
\hackcaption{The percentage of current that flows through the pinhole
against the side of the pinhole with the bottom layer held at 77K (main
graph) and 300K (inset).
The input volage is 0.3V.}
\label{p_77}
\end{figure}

There are several approaches to non-linear electro-thermal
modeling.\cite{Palisoc88,Georg}
Here, we alternate relaxation methods for the voltage
and the temperature at each node.
The total electrical current through each node is zero (Kirchhoff's law),
so that for a node $A$,
\begin{equation}
\label{Kirchhoff}
\sum_{i}\frac{(V_A-V_i)}{R_{Ai}} = 0\rlap{\quad,} 
\end{equation}
where $V_j$ is the voltage at node $j$ ($=A,i$)
and the sum runs over the nearest neighbors
of $A$.  The resistance
$R_{Ai}$ between nodes $A$ and $i$
depends
on the average of the temperatures $T_A$ and $T_i$ of the
two nodes and, in the case of the tunneling
layer, on the voltage difference $V_A-V_i$ through
Simmons's formula.\cite{simmons,Akerman03}%
\newbox\foobox\setbox\foobox\hbox{\onlinecite{Jonsson00}}
\bibhookdar{Akerman03}{Simmons gives $dI/dV$ at zero temperature.
We have incorporated heuristically an additional linear increase in
the overall conductance with temperature following the data in
Reference~\hbox{\unhbox\foobox}; a more detailed treatment would
take as its starting point thermal smearing of the Fermi
surface, as in }
Equation \eqref{Kirchhoff} is set
as a large sparse matrix and solved with the standard
\texttt{SLATEC} package.
Similarly, the total heat flux through a node is equal to the
heat generated Ohmically:
\begin{equation}
\label{thermal}
\sum_{i}\frac{(T_A-T_i)\,K_i(\frac12(T_A+T_i))}{X_{Ai}}
= \sum_{i}\frac{(V_A-V_i)^2}{(S_{Ai}R_{Ai})}\rlap{~,}
\end{equation}
where $K_i$ is the temperature-dependent thermal conductivity
corresponding to the material of node $i$, $X_{Ai}$ is the distance
between nodes $A$ and $i$, and $S_{Ai}$
is the cross-sectional area spanned by the link between the nodes.
First, \eqref{Kirchhoff} is solved; it must be iterated until
voltages (checked at the pinhole and away from the pinhole) converge, since
$R_{Ai}$ through the insulating layer depends on the voltage drop.
Then the temperature at each node is determined through \eqref{thermal},
and the procedure is repeated until voltages and temperatures both converge.

\section{Results}
We consider runs with the bottom layer held at 77K
and with the side of the pinhole ranging from 0.1$\,$nm to 1.5$\,$nm.
For an input voltage of 0.3V, we find the maximum steady-state temperature
at the center of the pinhole for the largest pinhole to be
86K.  This rise of 9K is comparable to that
estimated in Reference \onlinecite{Schmalhorst}.
The temperature rise rapidly falls off away from the pinhole.

As illustrated in Figure~\ref{d_c}, the differential
conductance may show positive or negative curvature, depending on
pinhole size; we graph normalized curves for the pinhole side ranging from
0.1$\,$nm to 1.5$\,$nm, with the separatrix between 1.0$\,$nm
and 1.1$\,$nm.
%
%
When its curvature is positive,
we can fit $dI/dV$ to Simmons's formula to
extract an effective barrier thickness and effective height.
These effective barrier parameters are
shown in Figures~\ref{e_t} and \ref{e_h}.
Without a pinhole, we recover essentially the input parameters.
As the pinhole grows in size, the fits remain quite good
(Figure~\ref{t_c}), but the effective barrier parameters
vary: the apparent barrier thickness decreases, while the apparent
barrier height increases.  The decreasing apparent thickness has the greatest
effect on the overall conductance (which increases), while
the increasing apparent height tends mostly to flatten the curve
of differential conductance.  Such trends are consistent with
a simple WKB treatment of tunneling.\cite{Quantum}
We obtain similar results with the bottom layer held at 300K;
the separatrix between positive and negative curvature occurs
for a pinhole side between 4.0$\,$nm and 4.1$\,$nm.

From the above simulations, we obtain the percentage of the total
current flowing through the pinhole 
(Figure~\ref{p_77}). 
This percentage
will increase with the side of the pinhole; the highest
value corresponding to a positive curvature of $dI/dV$
is 88\% at 77K. (The
highest value is 81\% at 300K.)

Our simulations  
support the contention that
a good fit of differential conductance to the Simmons form
fails, by itself, to verify the quality of a tunnel junction.

\begin{acknowledgments}	
We thank Johan {\AA}kerman for useful discussions.  DAR is
a Cottrell Scholar of Research Corporation, which has funded part of
this research.  Numerical work was carried out at the Research-Oriented
Computing Center of the University of South Florida.
\end{acknowledgments}

\bibliographystyle{modified}	
\bibliography{conductance0.bib}


{
\parskip1\baselineskip
\immediate\immediate\immediate\immediate\immediate\closeout\figcapwrite
\catcode`@11
\input\jobname-figcap.aux
\catcode`@12
}
\vfil\eject


\end{document}